\documentclass[letterpaper, 11pt]{article} 
\usepackage[top=1in, bottom=1in, left=1in, right=1in]{geometry}
\usepackage[utf8]{inputenc}
\usepackage[T1]{fontenc}
\usepackage{amsmath}
\usepackage{amsthm}
\usepackage{amsfonts}
\usepackage{amssymb}
\usepackage{url}
\usepackage{eucal}
\usepackage{mathrsfs}
\usepackage{float}
\usepackage[german,english]{babel}
\usepackage{subcaption}
\usepackage[T1]{fontenc}%
\usepackage{url}
\usepackage{epsfig}
\usepackage{epstopdf}
\usepackage{xcolor}

\newcommand{\beqa}{\begin{eqnarray}}
\newcommand{\eeqa}[1]{\label{#1}\end{eqnarray}}
\newcommand{\beq}{\begin{equation}}
\newcommand{\eeq}[1]{\label{#1}\end{equation}}

\newcommand{\bpm}{\begin{pmatrix}}
	\newcommand{\epm}{\end{pmatrix}}

\def\b0{\bf 0}

\title{On the effects of suitably designed space microstructures in the propagation of waves in time modulated composites}
\author{O. Mattei$^1$, V. Gulizzi$^2$}
\date{\small{$^1$ Department of Mathematics, San Francisco State University, CA 94132, USA,\\
		$^2$Department of Engineering, Universit{\'a} degli Studi di Palermo, Palermo, Italy. \\
		Emails: mattei@sfsu.edu, vincenzo.gulizzi@unipa.it}}
	
\begin{document}
\maketitle
\begin{abstract}
	\noindent
The amplitude of a pulse that propagates in a homogeneous material whose properties are instantaneously changed periodically in time will undergo an exponential increase, due to the interference between the reflected and transmitted pulses generated at each sudden switch. Here we resolve the issue by designing suitable reciprocal PT-symmetric space-time microstructures both in the one-dimensional and two-dimensional case, so that the interference between the scattered waves is such that the overall amplitude of the wave will be constant in time in each constituent material. Remarkably, for the geometries here proposed, a pulse will propagate with constant amplitude regardless of the impedance between the constituent materials, and for some, regardless of the wave speed mismatch. Given that the energy associated with the wave will increase exponentially in time, this creates the possibility to exploit the stable propagation of the pulse to accumulate energy for harvesting. 
\end{abstract}

Although the potential of time-modulated materials was first realized in the late Fifties (see \cite{Cullen:1958:TWP,Tien:1958:PAF,Morgenthaler:1958:VME,Louisell:1958:PAS,Honey:1960:WBU}, just to name a few),  they have experienced a widespread interest only in recent years, due to the associated extreme wave phenomena. Some examples are antireflection temporal coatings \cite{Pacheco-Pena:2020:AC}, temporal pumping in electromechanical waves \cite{Xia:2021:EOT}, nonreciprocal wave phenomena \cite{Nassar:2017:MPC}, temporal and spatiotemporal crystals \cite{Zhang:2017:ODT,Sharabi:2022:STP}, spatiotemporal cloaking \cite{McCall:2011:stc}, Bloch symmetry breaking \cite{Galiffi:2021:PLB}, and inverse prisms \cite{Akbarzadeh:2018:IVP}. 
See also the review articles \cite{Caloz:202:SPMI,Caloz:202:SPMII} and the comprehensive book \cite{Lurie:2017:ITD}. 

Here, for the sake of generality, we will not specify the type of wave phenomenon: we will consider any physics in which wave propagation can be described by a linear second-order wave equation of the form
\begin{equation}\label{homogeneous_wave_eq}
	\alpha \frac {\partial^2 u}{\partial x^2} = \beta\frac {\partial^2 u}{\partial t^2} \end{equation}
where \(\alpha\) and \(\beta\) are material parameters, defining the wave speed $c=\sqrt{\alpha/\beta}$ and the wave impedance $\gamma=\sqrt{\alpha\beta}$ of the material.

We are interested in the propagation of a pulse through a sequence of time interfaces, moments of time at which the material properties are instantaneously changed due to a sudden change in the applied field. The first experimental observation of  a time interface occurred in the context of water waves by instantaneously changing the effective gravity  \cite{Bacot:2016:TRH}. For mechanical waves, time interfaces are attained by using, for instance,  piezoelectric patches \cite{Marconi:2020:EON}. In electromagnetics, for which the realization of a time interface is more challenging, a photonic time interface was recently observed in a microwave transmission line \cite{Moussa:2022:OTR}. 

Each time the wave encounters a time interface, it will split into two waves traveling in opposite directions. Given that the amplitude of the wave as well as the momentum are conserved (see, e.g., \cite{Mendonca:2000:TPA,Lurie:2017:ITD}), the amplitudes of the two outgoing waves will be determined by the amplitude of the incoming wave, $u_i$, traveling from material 1 to material 2, by
\begin{equation}\label{ref_tran_time}
	u_-=\frac{\gamma_2-\gamma_1}{2\gamma_2}u_i,\quad u_+=\frac{\gamma_1+\gamma_2}{2\gamma_2}u_i
\end{equation}
$u_-$ being the wave traveling in the opposite direction and $u_+$ the one traveling in the same direction as the incoming wave.
As a result of time modulation, both the energy and the amplitude of the wave will increase exponentially in time (see, e.g., \cite{Lurie:2006:WPE,Lurie:2016:EAW,Torrent:2018:LST}). For instance, the amplitude of  the wavefront of  a pulse traveling through a time laminate will be given by
$$u_f=\left(\frac{\gamma_1+\gamma_2}{2}\right)^{n+m}\frac{1}{\gamma_1^n\gamma_2^m}u_i$$
where $u_i$ is the amplitude of the incoming pulse,  $n$  the number of switches from material 1 to material 2 and $m$ the ones  from material 2 to material 1. The goal of this letter is to introduce suitably designed space geometries to counteract the growth of the overall amplitude of the wave, due to time modulation. Indeed, if the spatial geometry is properly tailored, the scattering of the wave due to space interfaces can interfere with the scattering of the wave due to time interfaces to maintain the overall amplitude of the wave constant. We recall that, at a space interface, the amplitude and the flux of the wave are conserved, so that the amplitude of the reflected $u_r$ and transmitted $u_t$ waves as functions of the amplitude $u_i$ of the incoming wave will be
\begin{equation}\label{ref_tran_space}
	u_r=\frac{\gamma_1-\gamma_2}{\gamma_1+\gamma_2}u_i,\quad u_t=\frac{2\gamma_1}{\gamma_1+\gamma_2}u_i
\end{equation}
when the incoming wave travels from material 1 to material 2. 

Here, we will work with composites that are reciprocal and PT-symmetric, where P-symmetry refers to reflection invariance of the microstructure under the parity operation of spatial reflection, and T-symmetry stands for time reversal symmetry (note that for space-time composites, the two symmetries do not hold separately).  

Some examples of geometries achieving such an objective were proposed for the one-dimensional case in \cite{Mattei:2017:FPW}, within the theory of field patterns \cite{Mattei:2017:FP,Mattei:2017:FPA,Movchan:2022:FWT}:  the space and time interfaces are placed in such a way that the branching of the pulse due to the transmission and reflection at each interface will not cause a cascade of disturbances but rather a periodic pattern. Except for two remarkable one-dimensional examples of space-time checkerboards in which the materials have the same wave speed, all the examples proposed in \cite{Mattei:2017:FP,Mattei:2017:FPW,Mattei:2017:FPA} are such that, for certain values of the wave impedances, the condition of PT symmetry is unbroken \cite{Bender:1998:SSD}  and for others it is broken.  When the condition of  PT symmetry is unbroken, then the pulse propagates within the space-time material without any exponential growth in amplitude. The associated energy, however, still grows exponentially in time. In order to conserve energy, indeed, one needs, for instance, to introduce nonlinearity as showed in \cite{Deshmukh:2022:AEC}. In this letter, we first extend the results in \cite{Mattei:2017:FPW} by providing one-dimensional field-pattern materials for which the condition of PT-symmetry is always unbroken, regardless of the material properties. 

If the constituent materials have the same wave impedance, then, there will be no reflected wave when the pulse encounters either a space or a time interface (see equations \eqref{ref_tran_time} and \eqref{ref_tran_space}), and its amplitude will not increase in time (see, e.g., \cite{Lurie:2006:WPE,Lurie:2009:MAW}). 
In the case of a space-time composite with the two constituent materials having different impedances, a sufficient condition to ensure that a pulse will maintain a constant amplitude as it propagates in the material, is the following: the pulse has to encounter a space interface followed by a time interface between the same materials periodically.
Geometrically, this means that any characteristic line in a space-time diagram will need to cross a horizontal interface, after crossing a vertical interface.  In order for such a design principle to be satisfied by both the wavefront and the scattered waves, one has to choose a field pattern material \cite{Mattei:2017:FP,Mattei:2017:FPW,Mattei:2017:FPA,Movchan:2022:FWT}, for which the network of characteristic lines is locally periodic so that the scattered waves interact according to a precise pattern.

\begin{figure}[h!]
	\centering
	\includegraphics[width=0.8\textwidth]{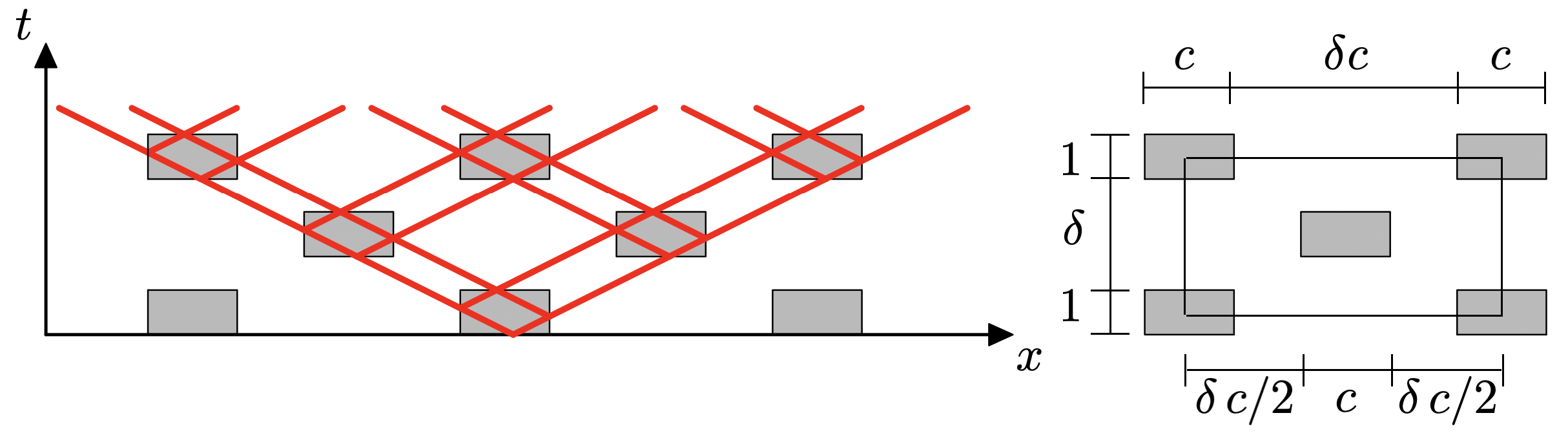}
	\caption{A field pattern material where the two materials in white and gray have the same wave speed $c$.  }
	\label{fig_Same_speed_geo_1d}
\end{figure}

\begin{figure}[h!]
	\centering
	\includegraphics[width=0.95\textwidth]{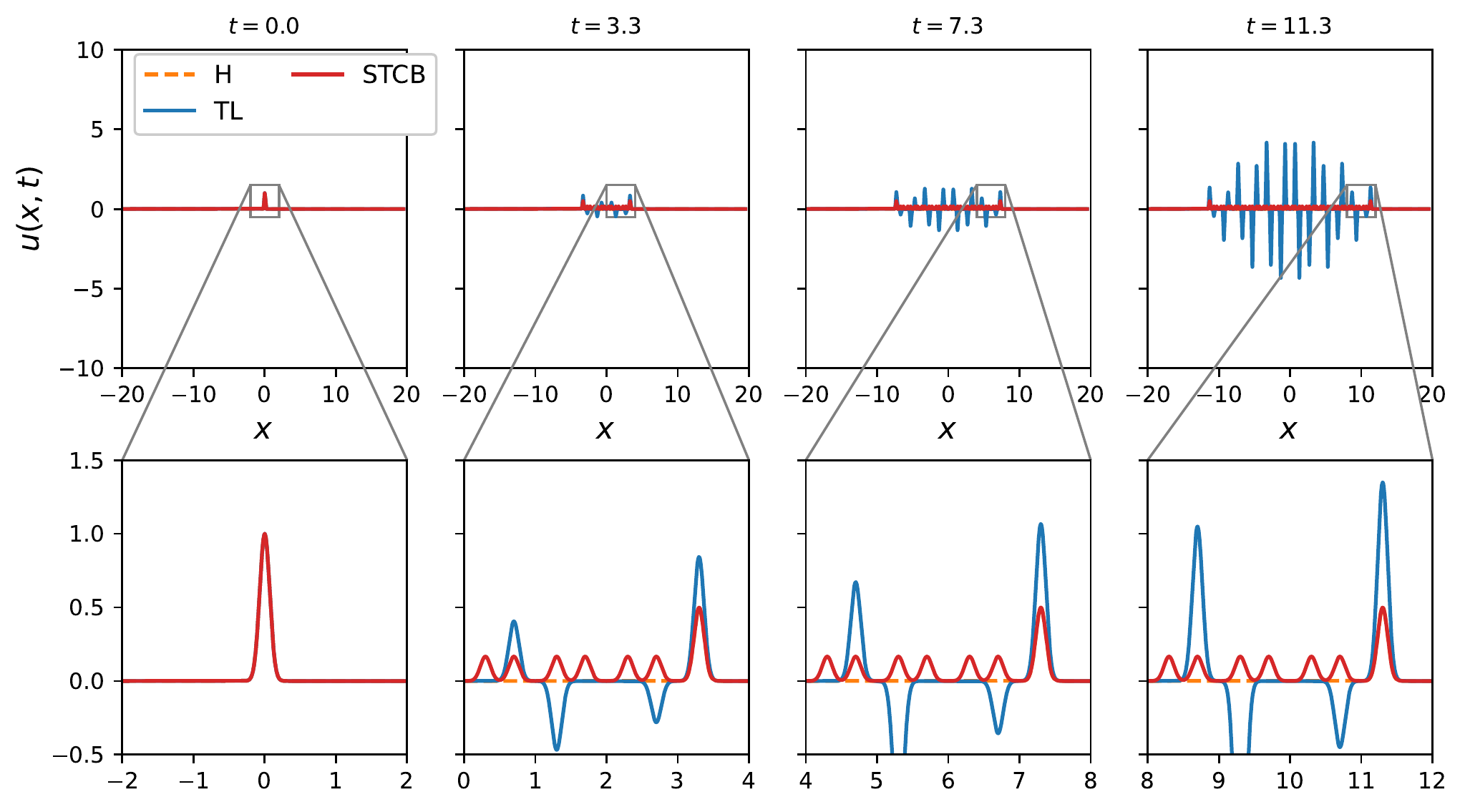}
	\caption{Snapshots of the amplitude of a Gaussian pulse of the form $u(x,0)=\exp(-100x^2)$, as it propagates through a homogeneous material with $\alpha=\beta=1$ (orange), a time laminate with  $\alpha_1=\beta_1=1$, $\alpha_2=\beta_2=0.5$, and period of modulation $T=1$ (blue), the space-time geometry in figure \ref{fig_Same_speed_geo_1d}, with $\delta=1$, $\alpha_1=\beta_1=1$ and $\alpha_2=\beta_2=0.5$ (red). }
	\label{comparison_intensity_1d}
\end{figure}

\begin{figure}[h!]
	\centering
	\includegraphics[width=0.5\textwidth]{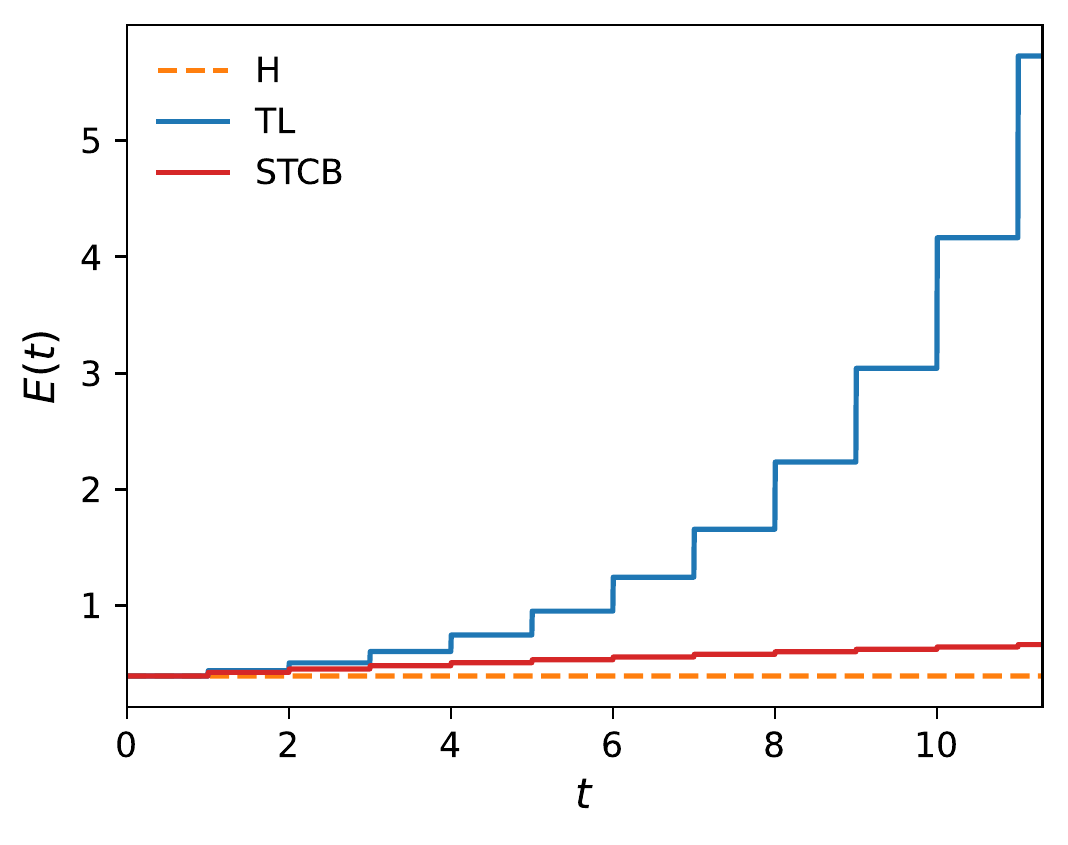}
	\caption{Energy associated with the wave propagation described in figure \ref{comparison_intensity_1d}.}
	\label{comparison_energy_1d}
\end{figure}

The only two field-pattern materials proposed in the literature (see \cite{Mattei:2017:FP,Mattei:2017:FPW,Mattei:2017:FPA,Movchan:2022:FWT}) for which a pulse propagates at constant amplitude, regardless of the impedance mismatch, are space-time checkerboards in which the constituent materials have the same wave speed. A generalization of such a geometry  is the one showed in figure \ref{fig_Same_speed_geo_1d},  where $\delta\in[0,+\infty)$. The right-going and left-going  wavefronts will always have amplitude equal to the initial one when traveling through material 1 (gray), and will have amplitude equal to $2\gamma_1/(\gamma_1+\gamma_2)$ when traveling through material 2 (white), regardless of the impedance mismatch: the smaller the impedance mismatch, the smaller the difference in amplitude. The scattered waves will interfere such that the oscillations in the wake of the wave will always have the same amplitude, which could be found analytically, as also showed in figure \ref{comparison_intensity_1d} (the numerical simulations are performed by using the algorithm in \cite{Gulizzi:2022:MWP}). The associated energy is depicted in red in figure \ref{comparison_energy_1d}.

\begin{figure}[h!]
	\centering
	\includegraphics[width=0.9\textwidth]{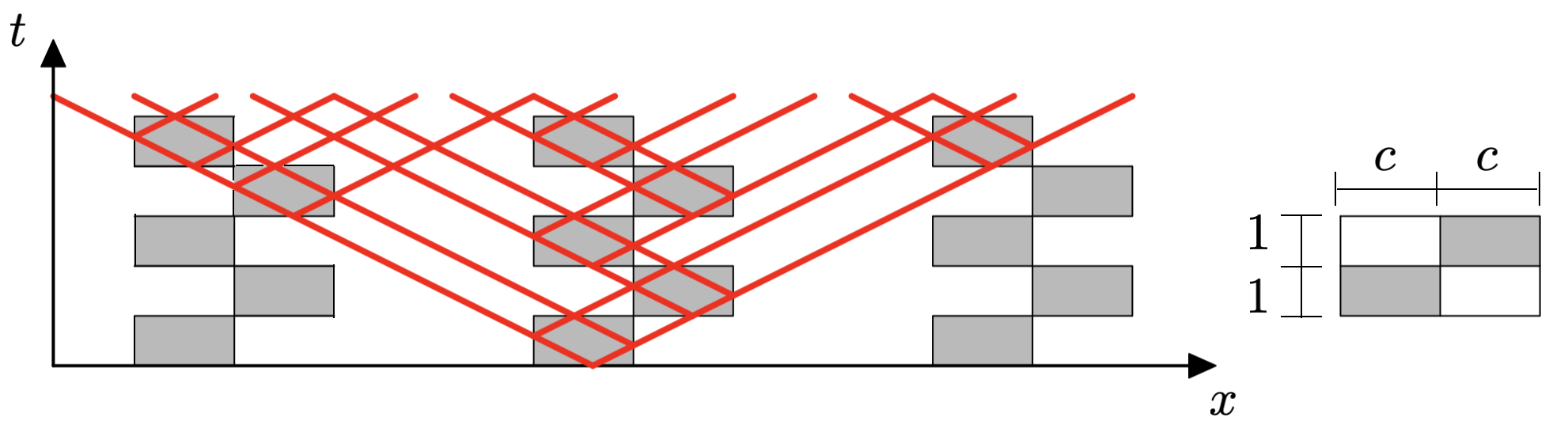}
	\caption{A field pattern material where the two materials in white and gray have the same wave speed $c$.  }
	\label{fig_Same_spee_non_rec}
\end{figure}

Another example of a field-pattern material with the two components having the same speed and satisfying our design principle is represented in figure \ref{fig_Same_spee_non_rec}. 

\begin{figure}[h!]
	\centering
	\includegraphics[width=0.9\textwidth]{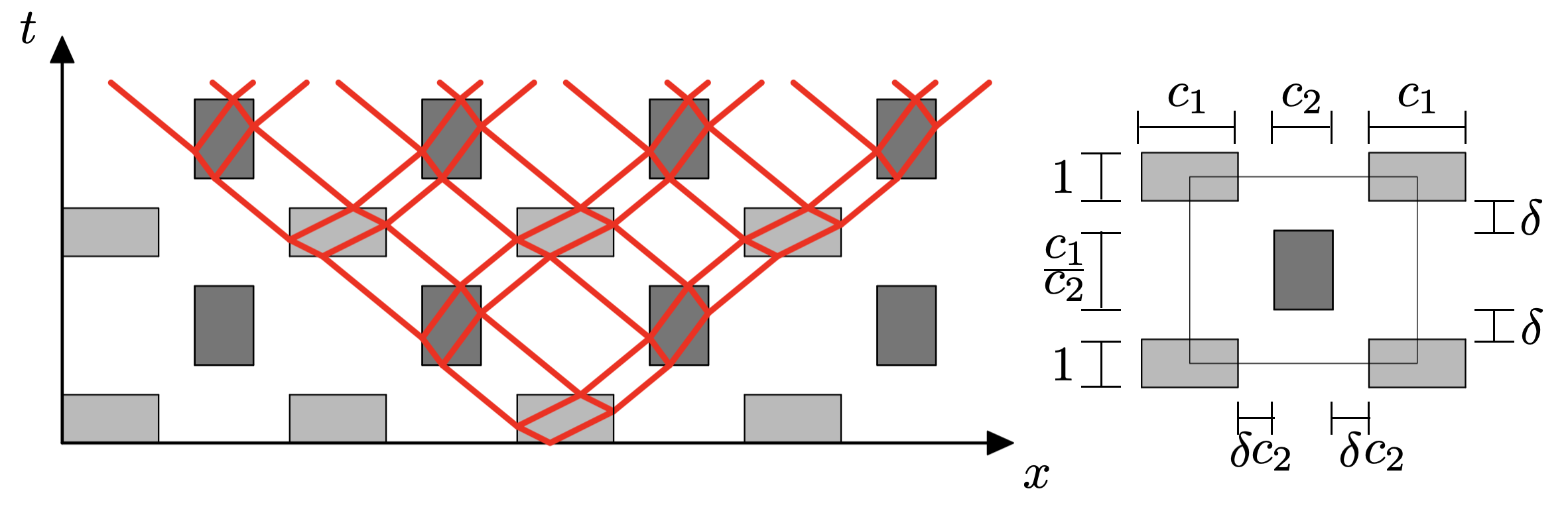}
	\caption{A field pattern material where the gray material has wave speed $c_1$, the white material $c_2$, and the dark gray $c_3=c_1^2/c_2$.  }
	\label{fig_Gen_1d}
\end{figure}

A space-time geometry  for which the wave propagates with constant amplitude regardless of both the wave impedance mismatch and the wave speed mismatch is the one depicted in figure \ref{fig_Gen_1d}, with \(\delta\in[0,\infty)\).
The wave speed of the gray and white materials, $c_1$ and $c_2$, respectively, is arbitrary and the one of the dark gray material is given by $c_3= c_1^2/c_2$. 
If the initial amplitude of the right or left wavefront is $u_i$, then it will have amplitude equal to $u_i$ in material 1 (gray), \(2\gamma_1u_i/(\gamma_1+\gamma_2)\) in material 2 (white), and  \(\gamma_1(\gamma_2+\gamma_3)u_i/(\gamma_2(\gamma_1+\gamma_2))\) in material 3 (dark gray). 
\begin{figure}[h!]
	\centering
	\includegraphics[width=0.95\textwidth]{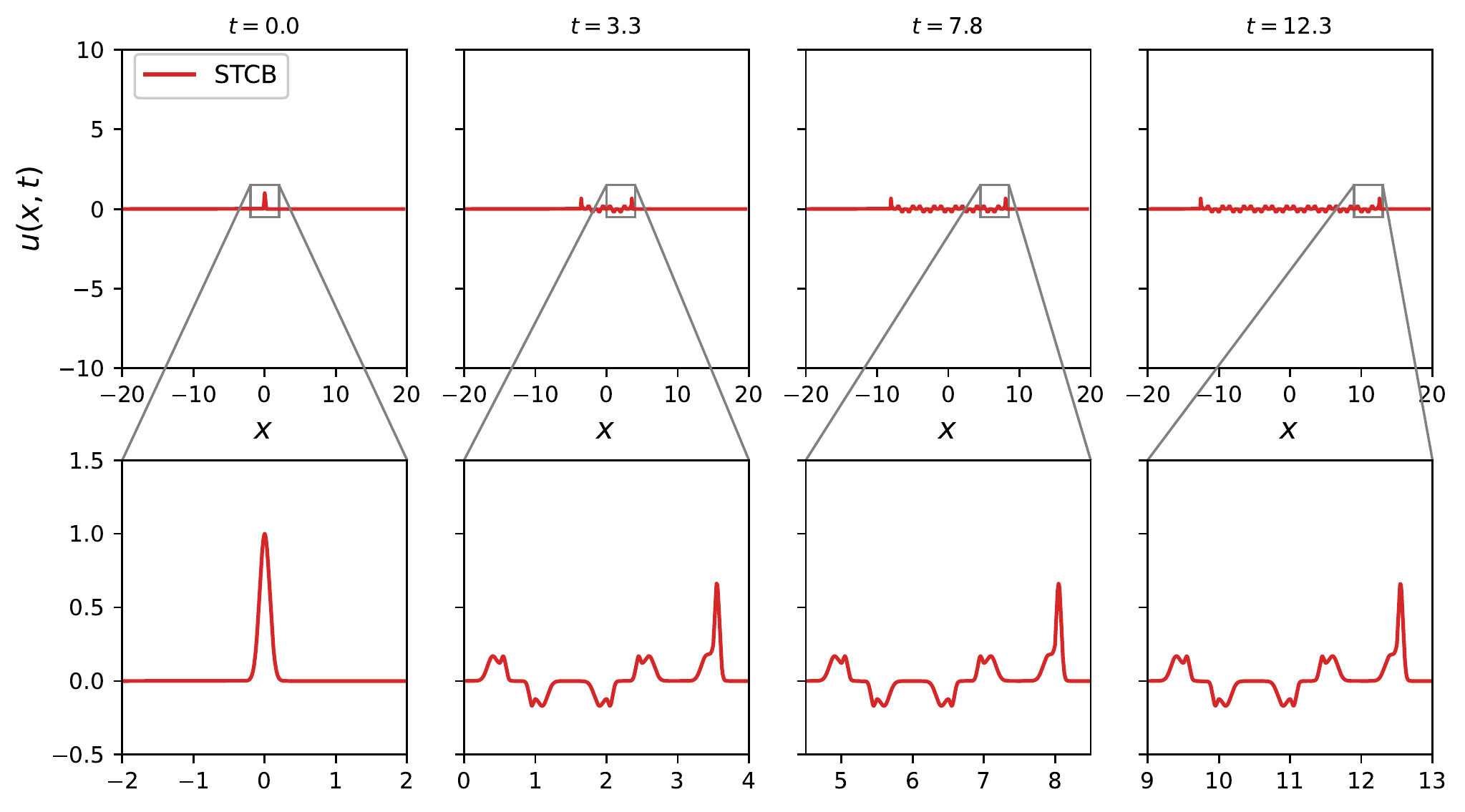}
	\caption{Snapshots of the amplitude of a Gaussian pulse of the form $u(x,0)=\exp(-100x^2)$, as it propagates through the space-time geometry in figure \ref{fig_Gen_1d}, with $\delta=0$, $\alpha_1=0.25$, $\beta_1=1$, $\alpha_2=\beta_2=1$, $\alpha_3=1$ and $\beta_3=4$.  }
	\label{comparison_intensity_1d_general}
\end{figure}
\begin{figure}[h!]
	\centering
	\includegraphics[width=0.5\textwidth]{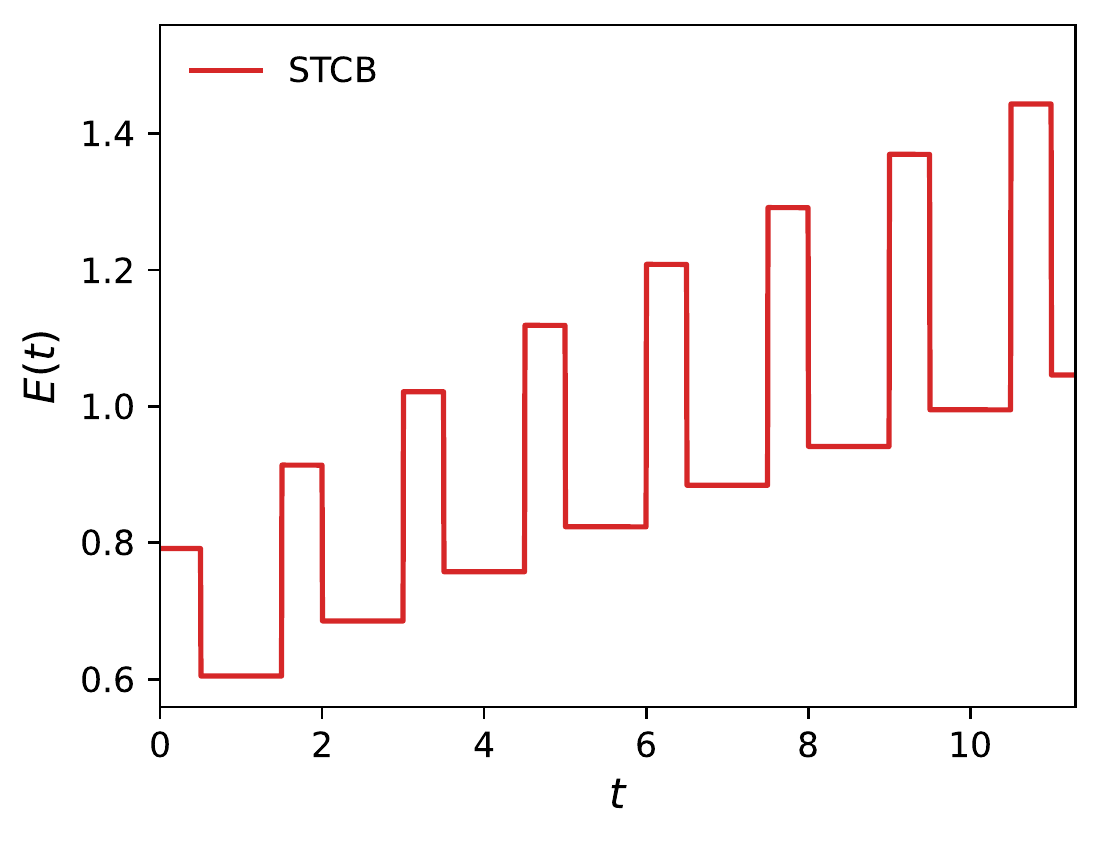}
	\caption{Energy associated with the wave propagation described in figure \ref{comparison_intensity_1d_general}.}
	\label{comparison_energy_1d_general}
\end{figure}
This is the most general design that ensures propagation of the wave at constant amplitude, regardless of the impedance mismatch.

\begin{figure}[h!]
	\centering
	\begin{subfigure}{.45\textwidth}
		\includegraphics[width=0.95\textwidth]{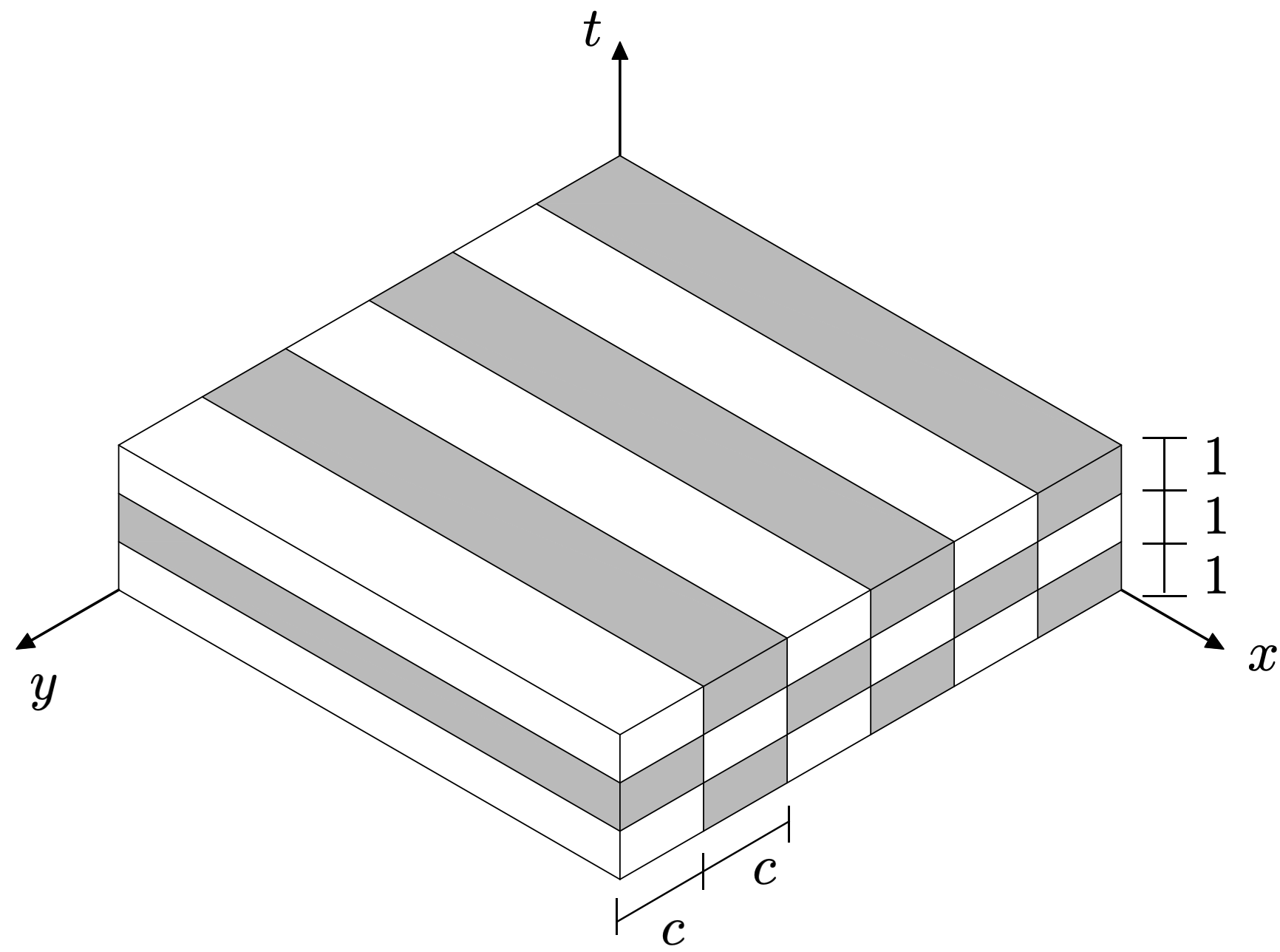}
		\caption{}	\label{fig_lam_check_2d}
	\end{subfigure}
	\begin{subfigure}{.45\textwidth}
		\includegraphics[width=0.95\textwidth]{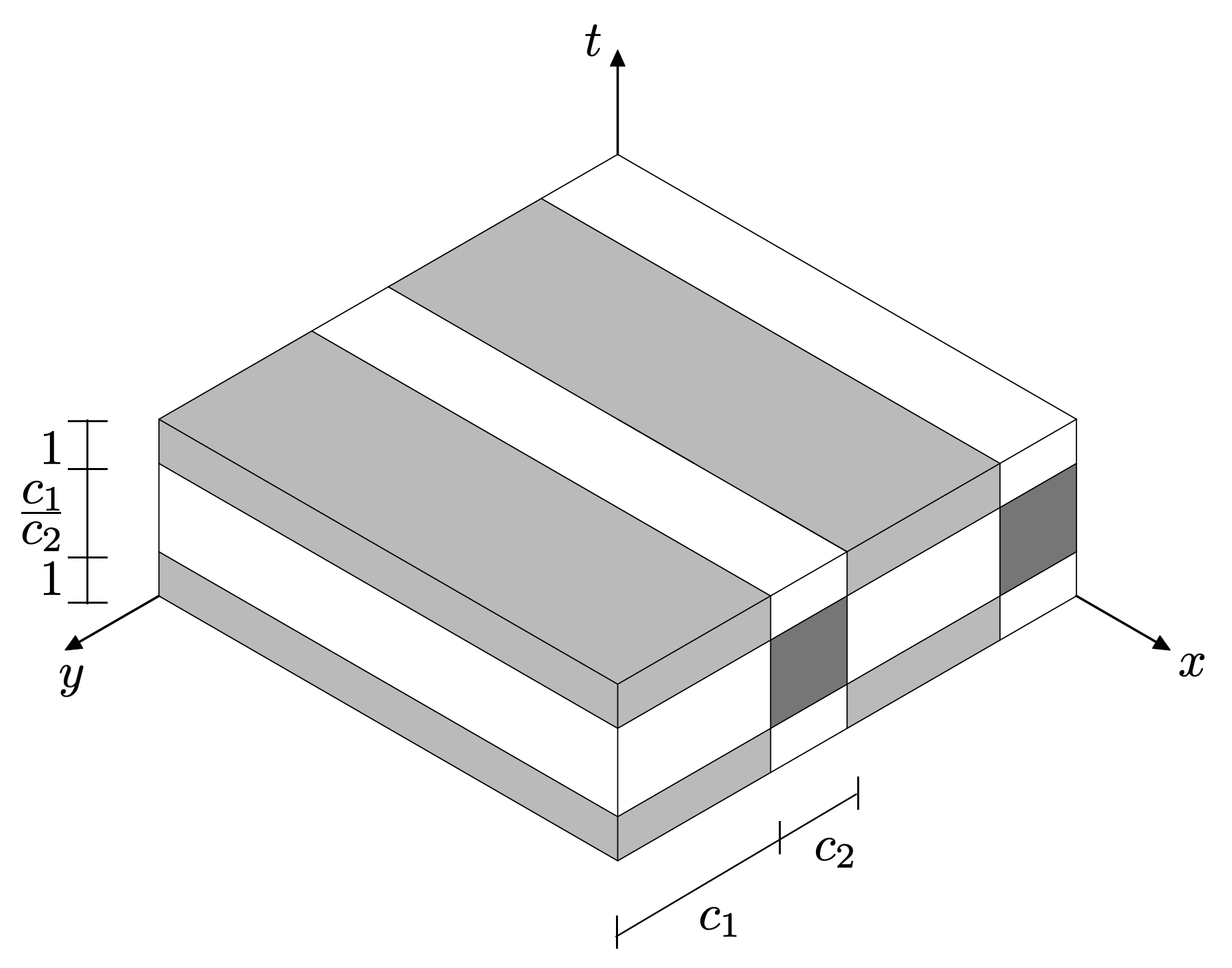}
		\caption{}
		\label{fig_2d_diff_speed}
	\end{subfigure}
	\caption{(a) A field pattern material where the two materials in white and gray have the same wave speed $c$.   This is the two-dimensional extension of the geometry of figure  \ref{fig_Same_speed_geo_1d} when $\delta=0$. (b) A field pattern material where the gray material has wave speed $c_1$, the white material $c_2$, and the dark gray $c_3=c_1^2/c_2$. This is the two-dimensional extension of geometry of figure  \ref{fig_Gen_1d} when $\delta=0$. }
\end{figure}

\begin{figure}
	\centering
	\includegraphics[width = 0.9\textwidth]{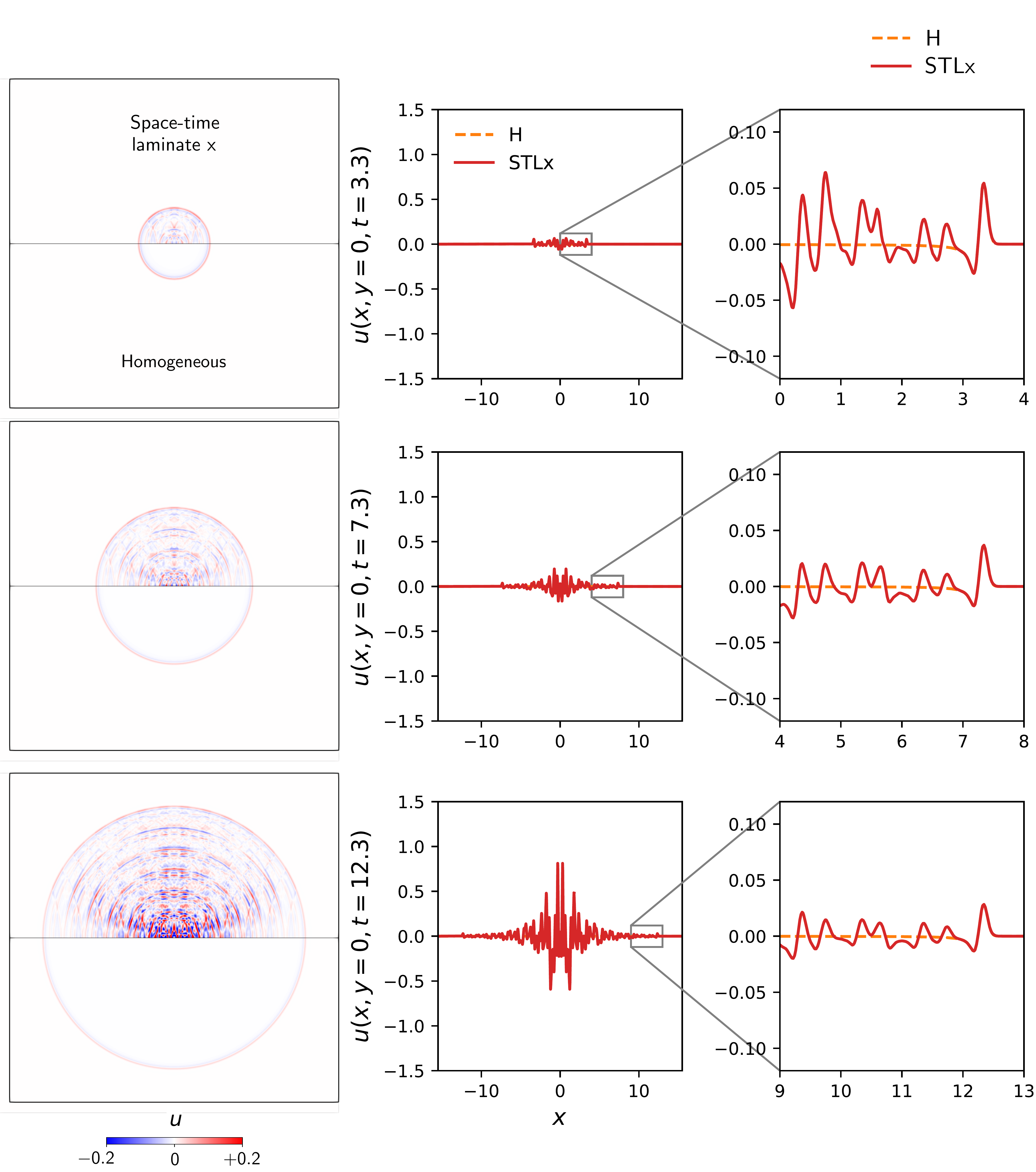}
	\caption
	{Amplitude of a Gaussian pulse of the form $u(x,y,0)= \exp(-100(x^2+y^2))$, as it propagates through a homogeneous material with $\alpha=\beta=1$ (orange), and  the space-time geometry in \ref{fig_lam_check_2d}, with $\alpha_1=\beta_1=1$ and $\alpha_2=\beta_2=0.5$ (red). 
	}
	\label{fig:11-time_history-2d}
\end{figure}

\begin{figure}[h!]
	\centering
	\begin{subfigure}{0.45\textwidth}
		\includegraphics[width=\textwidth]{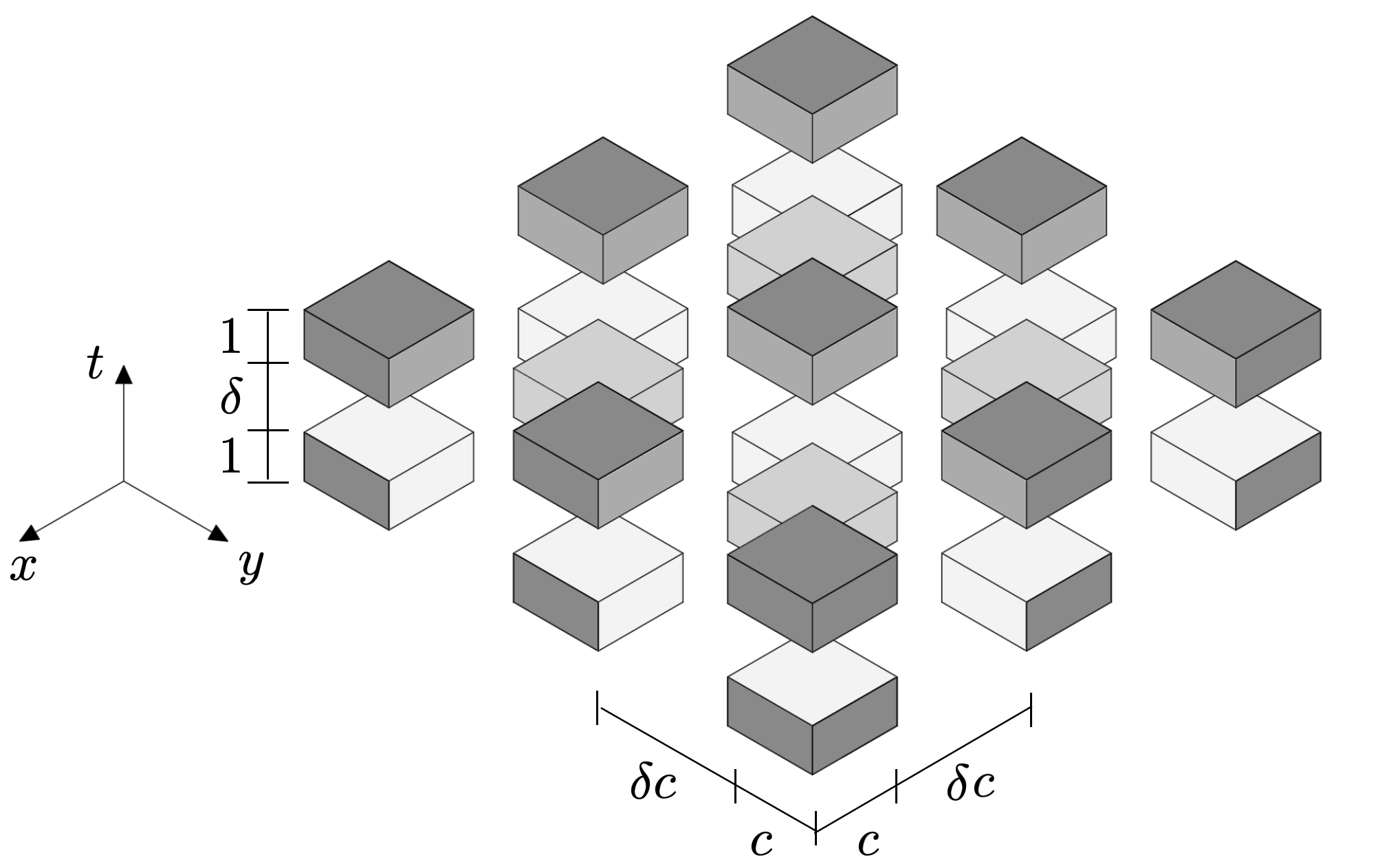}
		\caption{}
		\label{fig_Gen_2d_same_speeds}
	\end{subfigure}
	\begin{subfigure}{0.45\textwidth}
		\includegraphics[width=\textwidth]{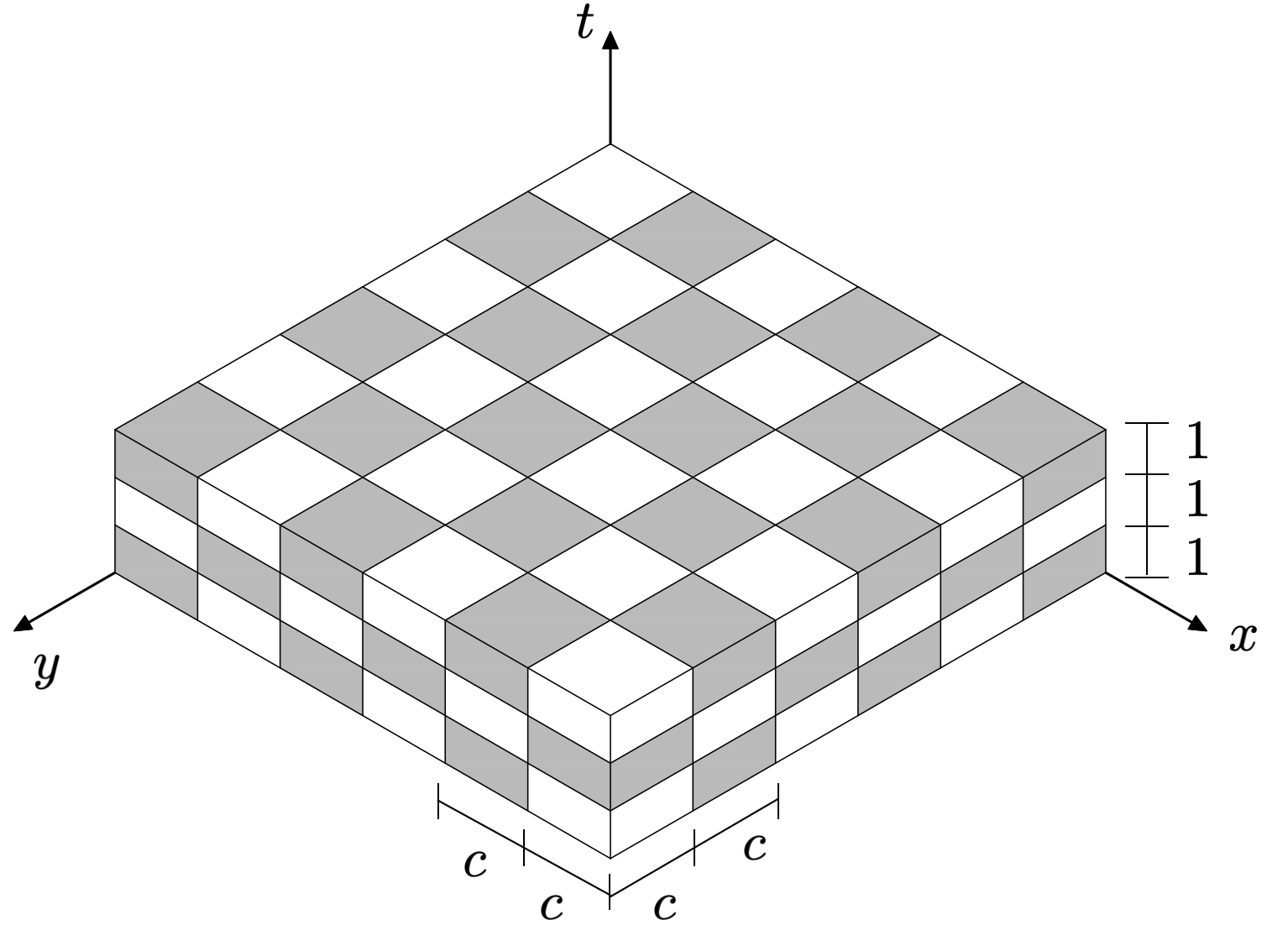}
		\caption{}
		\label{fig_2d_checker}
	\end{subfigure}
	\caption{(a) A two-dimensional field pattern material where the matrix and the inclusions have the same wave speed $c$, and $\delta\in[0,\infty)$. The different shades of gray are only to emphasize the depth of the three-dimensional  geometry. (b) The geometry in (a) when $\delta=0$.}	
\end{figure}

\begin{figure}
	\centering
	\includegraphics[width = 0.9\textwidth]{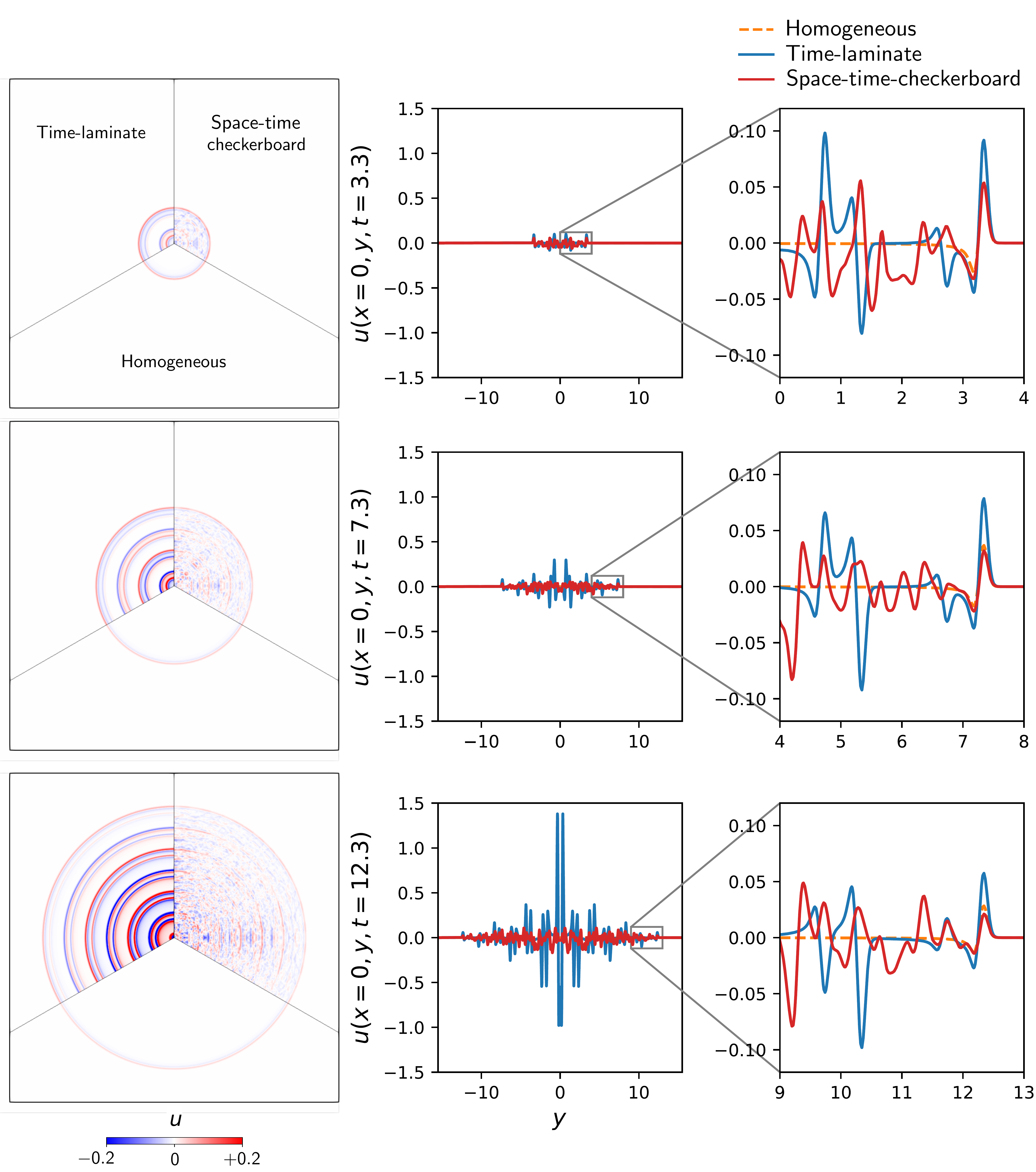}
	\caption
	{Amplitude of a Gaussian pulse of the form $u(x,y,0)= \exp(-100(x^2+y^2))$, as it propagates through a homogeneous material with $\alpha=\beta=1$ (orange), a time laminate with $\alpha_1=\beta_1=1$, $\alpha_2=\beta_2=0.5$, and modulation period $T=1$ (blue), and the space-time geometry in figure \ref{fig_2d_checker}, with $\alpha_1=\beta_1=1$ and $\alpha_2=\beta_2=0.5$ (red). 
	}
	\label{fig:3-time_history-2d}
\end{figure}

\begin{figure}
	\centering
	\includegraphics[width = 0.5\textwidth]{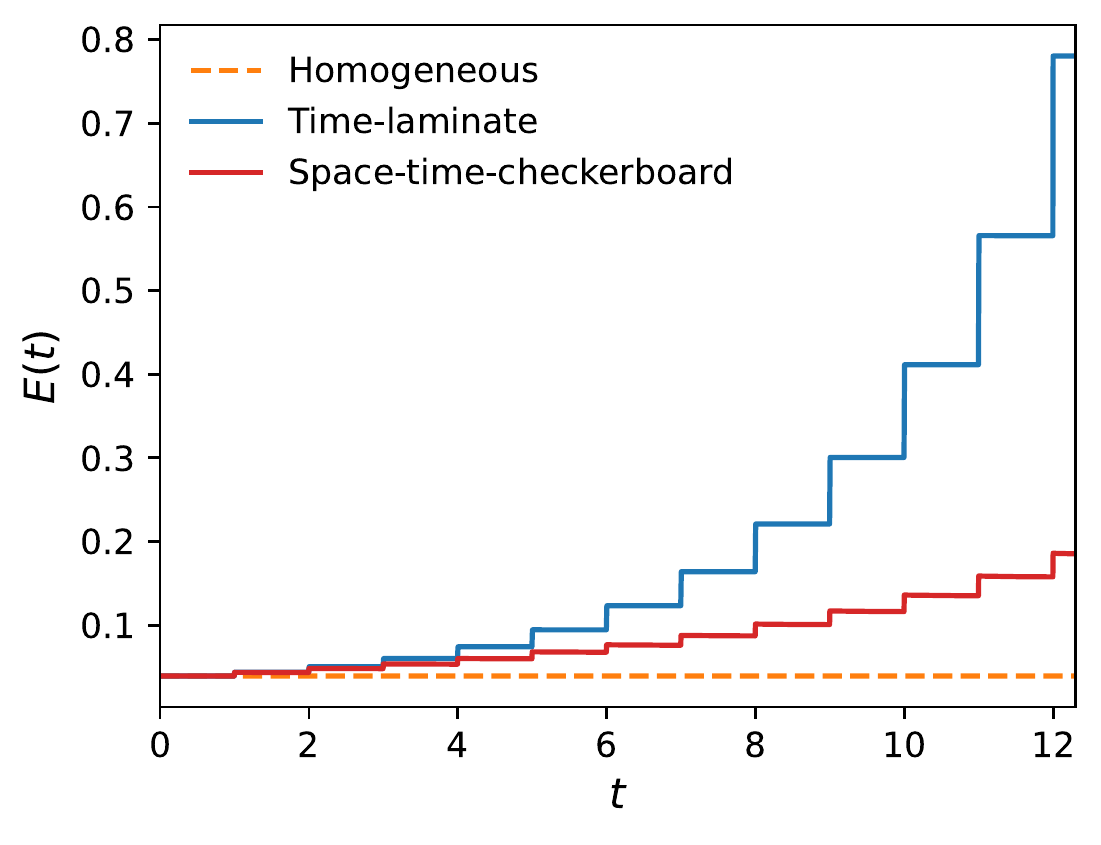}
	\caption
	{Energy associated with the wave propagation described in figure \ref {fig:3-time_history-2d}}
	\label{fig:4-total_energy-2d}
\end{figure}

In the two-dimensional case, if the pulse is a unidirectional Gaussian, $u(x,y,0)=A\exp(-By^2)$, then, the results obtained in the one-dimensional case trivially hold, when the space-time geometries are the ones in figures \ref{fig_lam_check_2d} and \ref{fig_2d_diff_speed}. 
Propagation of a symmetrical Gaussian, $u(x,y,0)= A\exp(-B(x^2+y^2))$, in either one of the space-time geometries illustrated in figures \ref{fig_lam_check_2d} and \ref{fig_2d_diff_speed}, will result in a wavefront with constant amplitude in each material and a wake that increases amplitude in time, due to the spatial asymmetry of the geometry, as showed in figure \ref{fig:11-time_history-2d}. 
However, the wake will have constant amplitude if one considers a symmetric space-time geometry like the one illustrated in figures \ref{fig_Gen_2d_same_speeds} and \ref{fig_2d_checker}, see figure \ref{fig:3-time_history-2d}. Figure  \ref{fig:4-total_energy-2d} shows the associated energy. 

These are the first two-dimensional field-pattern materials ever proposed.  Given that the condition of PT-symmetry is always unbroken, they are promising candidates to create energy harvesting devices: as the wave travels without any instability in the material, the associated energy increases in time and can be harvested. 


\begin{figure}
	\centering
	\includegraphics[width = \textwidth]{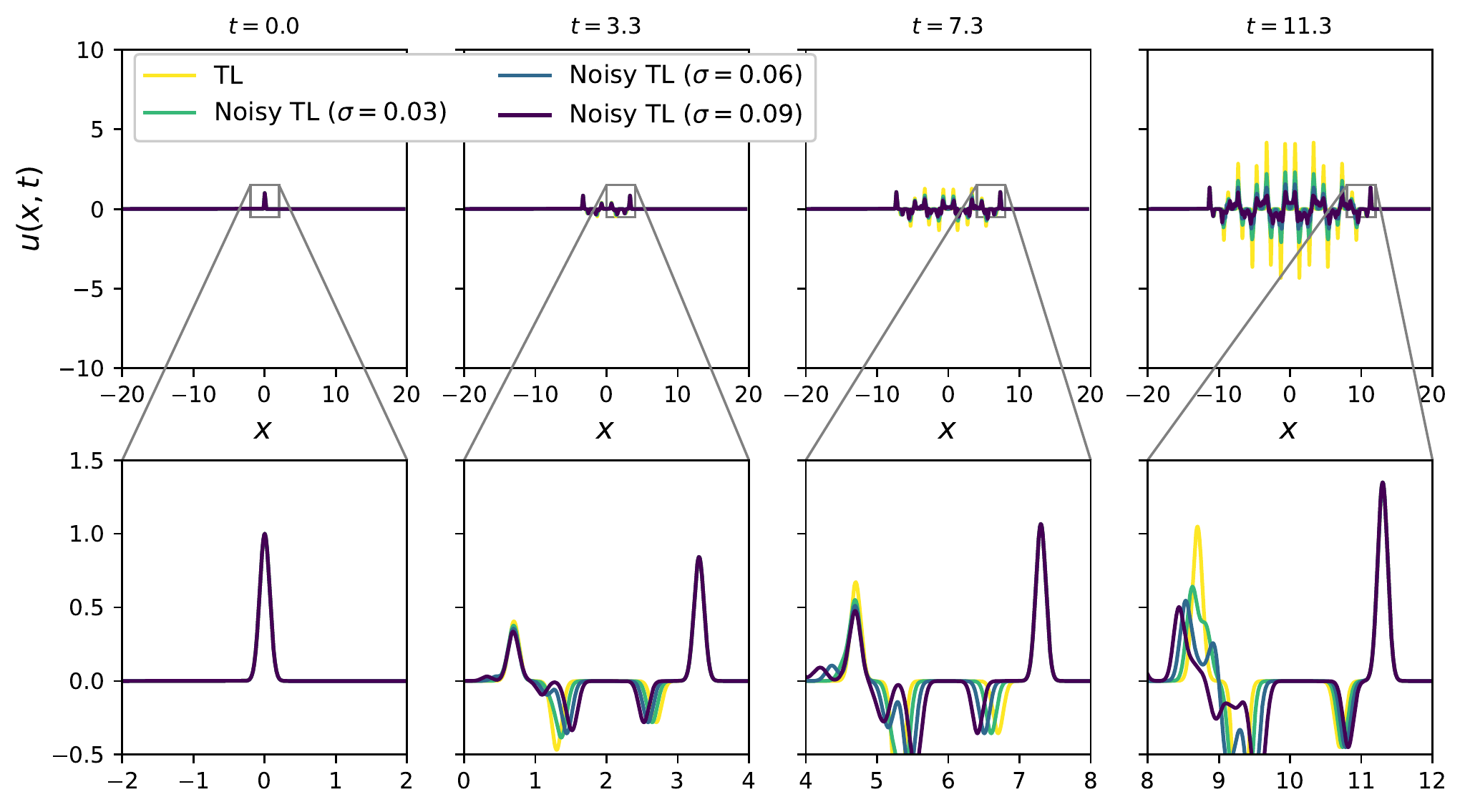}
	\caption
	{Snapshots of the amplitude of a Gaussian pulse of the form $u(x,0)=\exp(-100x^2)$, as it propagates through a time laminate, with $\alpha_1=\beta_1=1$, $\alpha_2=\beta_2=0.5$, and the period of time modulation $T=1$, with various degrees of disorder. }
	\label{fig:noise_time_history-1d_laminate}
	
\end{figure}

\begin{figure}
	\centering
	\includegraphics[width = 0.5\textwidth]{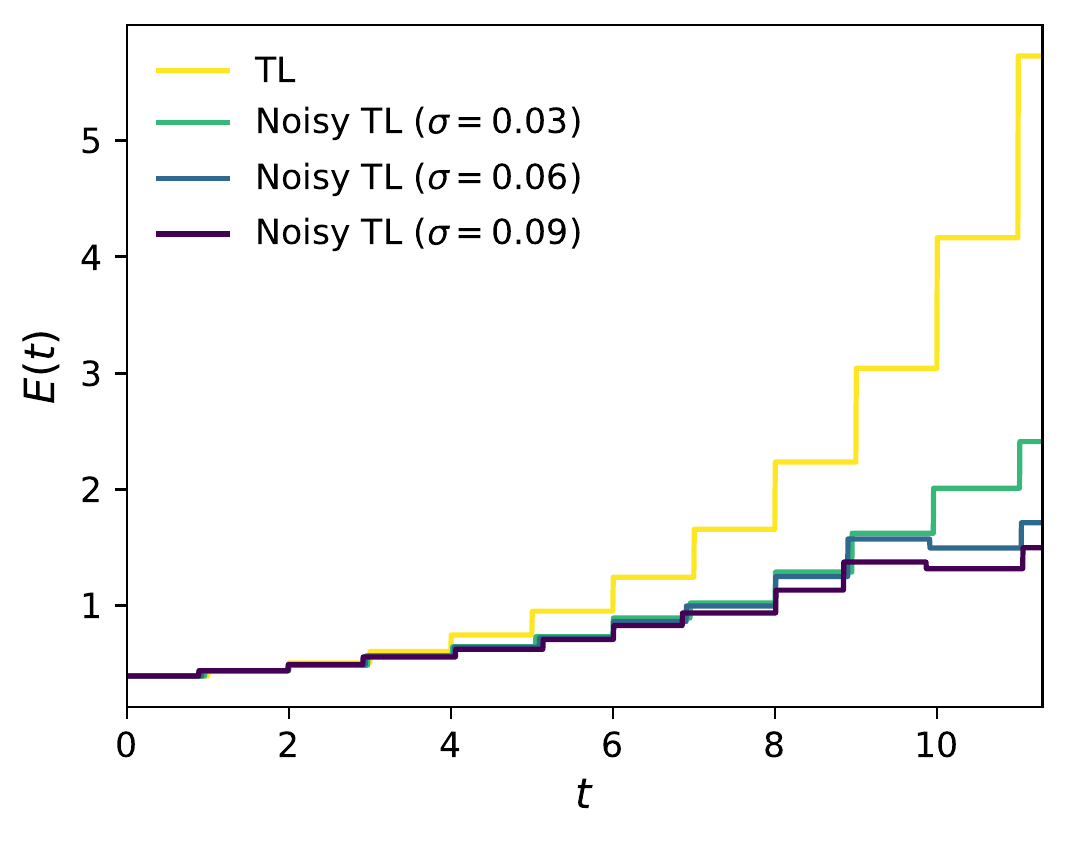}
	\caption
	{	The energy associated with the wave propagation described in figure \ref {fig:noise_time_history-1d_laminate}.}
	\label{fig:noise_energy-1d_laminate}
\end{figure}

\begin{figure}
	\centering
	\includegraphics[width = \textwidth]{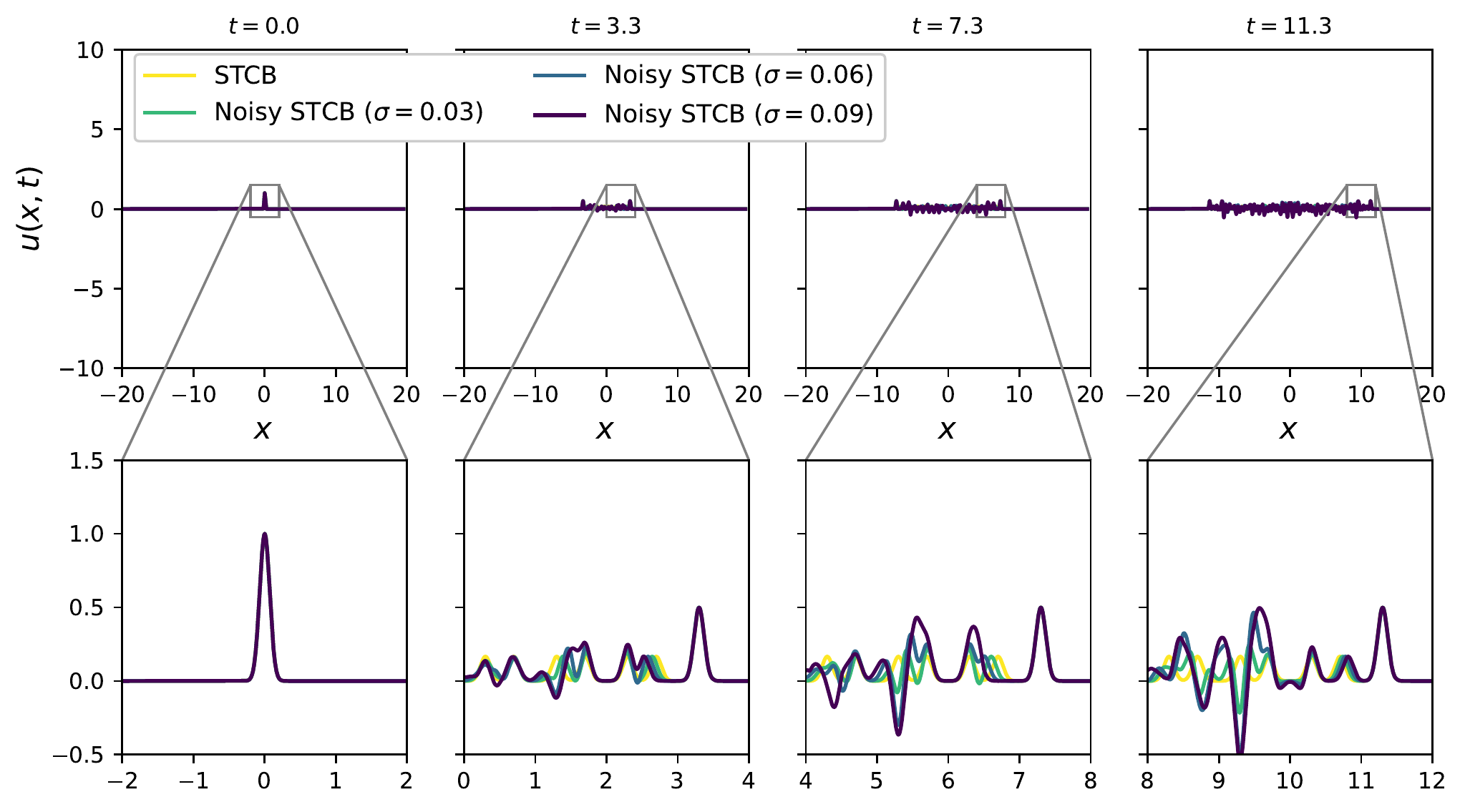}
	\caption
	{Snapshots of the amplitude of a Gaussian pulse of the form $u(x,0)=\exp(-100x^2)$, as it propagates through the space-time geometry of figure \ref{fig_Same_speed_geo_1d}, with $\alpha_1=\beta_1=1$ and $\alpha_2=\beta_2=0.5$, with various degrees of disorder.  }
	\label{fig:noise_time_history-1d_check}
\end{figure}

\begin{figure}
	\centering
	\includegraphics[width = 0.5\textwidth]{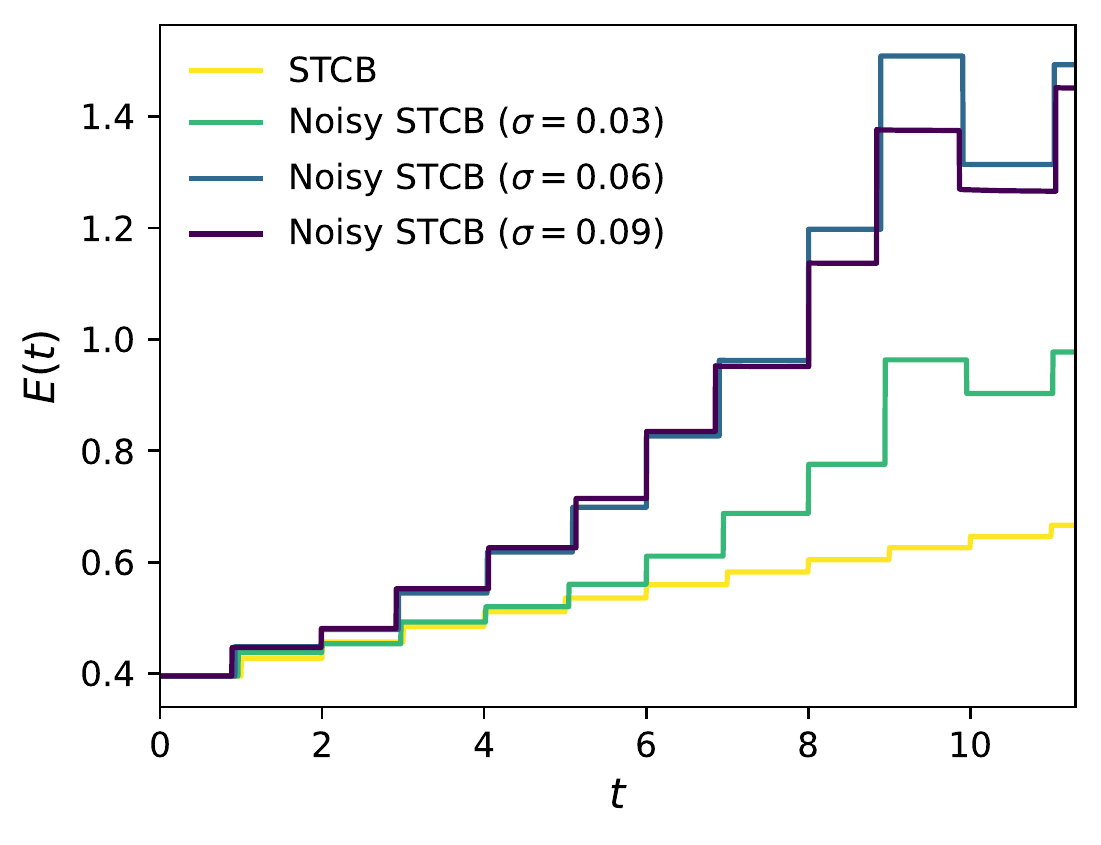}
	\caption
	{The energy associated with the wave propagation described in figure \ref {fig:noise_time_history-1d_check}.}
	\label{fig:noise_energy-1d_check}
\end{figure}

Finally, given that the constant overall amplitude of the wave is guaranteed only if the time interfaces are applied at specific moments of time, we study the effects of noise. Following \cite{Apffel:2022:EIW}, we consider the time interfaces occurring at random times $T_n=n+\epsilon_n$, with $n$ integer and $\epsilon_n$ chosen independently and uniformly in $[-\sqrt{3}\sigma,\sqrt{3}\sigma]$, with $\sigma$ being the noise standard deviation. As showed in figures \ref{fig:noise_time_history-1d_laminate} and \ref{fig:noise_energy-1d_laminate} for a time laminate, the bigger the disorder, the smaller the increase in the amplitude of the wave and energy, which is in agreement with \cite{Apffel:2022:EIW}. The opposite occurs when disorder is introduced in the space-time geometry illustrated in figure \ref{fig_Same_speed_geo_1d}, see figures \ref{fig:noise_time_history-1d_check} and \ref{fig:noise_energy-1d_check}, which suggests one has to reduce the noise to a minimum, in order to maintain the overall amplitude of the wave constant and the growth of the energy slow. 

\section*{Acknowledgments}
OM thanks the National Science Foundation for support through
grant DMS-2008105. 
\bibliographystyle{abbrv}
\bibliography{newref.bib,tcbook.bib,refer.bib}
\end{document}